\documentclass[sigconf]{acmart}

\usepackage{todonotes}
\usepackage{url}
\usepackage{hyperref}
\usepackage{booktabs}
\usepackage{multirow}
\usepackage{amssymb}
\usepackage{url}
\usepackage{flushend}

\newtheorem{theorem}{Theorem}
\newtheorem{hypothesis}[theorem]{Hypothesis}

\def\CVSS{{\texttt {CVSS}}}

\def\AV{{\texttt {AV}}}
\def\AC{{\texttt {AC}}}
\def\UI{{\texttt {UI}}}
\def\PR{{\texttt {PR}}}

\def\C{{\texttt {C}}}
\def\I{{\texttt {I}}}
\def\A{{\texttt {A}}}
\def\CI{{\texttt {C}}}
\def\II{{\texttt {I}}}
\def\AI{{\texttt {A}}}

\graphicspath{{../Figures/}}

\begin{document}
%
\title{Identifying Relevant Information Cues for Vulnerability Assessment Using \CVSS}


\author{Luca Allodi}
\orcid{0000-0003-1600-0868}
\affiliation{%
  \institution{Eindhoven University of Technology}
  \streetaddress{De Zaale, Eindhoven, The Netherlands, 5600 MB}
}
\email{l.allodi@tue.nl}

\author{Sebastian Banescu\\Henning Femmer}
\affiliation{%
  \institution{Munich Technical University}
  \streetaddress{Munich, DE}
  }
\email{name.surname@tum.de}

\author{Kristian Beckers}
\affiliation{%
  \institution{Social Engineering Academy (SEA) GmbH}
  \streetaddress{Frankfurt am Main, DE}
}
\email{kristian.beckers@social-engineering.academy}

\renewcommand{\shortauthors}{L. Allodi et al.}

\begin{abstract}
The assessment of new vulnerabilities is an activity that accounts for information from several data sources and produces a `severity' score for the vulnerability. The Common Vulnerability Scoring System (\CVSS) is the reference standard for this assessment. Yet, no guidance currently exists on \emph{which information} aids a correct assessment and should therefore be considered. 
 In this paper we address this problem by evaluating which information cues increase (or decrease) assessment accuracy.
  We devise a block design experiment with 67 software engineering students with varying vulnerability information and measure scoring accuracy under different information sets.
 We find that baseline vulnerability descriptions provided by standard vulnerability sources provide only part of the information needed to achieve an accurate vulnerability assessment. Further, we find that  additional information on \texttt{assets}, \texttt{attacks}, and \texttt{vulnerability type} contributes in increasing the accuracy of the assessment; conversely, information on \texttt{known threats} misleads the assessor and decreases assessment accuracy and should be avoided when
 assessing vulnerabilities.
 These results go in the direction of formalizing the vulnerability communication to, for example, fully automate security assessments.
\end{abstract}



\copyrightyear{2018} 
\acmYear{2018} 
\setcopyright{acmcopyright}
\acmConference[CODASPY '18]{Eighth ACM Conference on Data and Application Security and Privacy}{March 19--21, 2018}{Tempe, AZ, USA}
\acmBooktitle{CODASPY '18: Eighth ACM Conference on Data and Application Security and Privacy, March 19--21, 2018, Tempe, AZ, USA}
\acmPrice{15.00}
\acmDOI{10.1145/3176258.3176340}
\acmISBN{978-1-4503-5632-9/18/03}
\fancyhead{}

\maketitle

\section{Introduction}
\label{sec:intro}




Addressing software vulnerabilities is an important process in any software development project~\cite{HOWARD-06-SDL} to maintain software quality and mitigate risk of attack for the users. Several standards, such as PCI-DSS for the management of credit card information and NIST's SCAP protocol (adopted for example by the U.S. DoD directive 8500.01), require the use of the Common Vulnerability Scoring System~\cite{first-2015-cvss3}, \CVSS, as the metric of choice for vulnerability measurement and prioritisation~\cite{PCI-DSS-DOC,Scarfone-2010-SCAP}.
The \CVSS\ specification~\cite{first-2015-cvss3} describes a framework that the assessor follows to transform information about the vulnerability into a \CVSS\ score, and provides a number of `dimensions' or `metrics' over which the assessor performs his or her evaluation. For example, the assessor may evaluate that the vulnerability can be remotely accessed, and assign a \texttt{Network} value to the \CVSS\ metric \texttt{Attack Vector}; similarly, he or she may conclude that a successful attack requires the victim user to perform specific actions for the attack to be successful, and assign a \texttt{Required} value to the \CVSS\ metric \texttt{User Interaction}.

The result of these assessments depends strongly on what information on the vulnerability is available to the assessor. Notably, this information may vary substantially, ranging from general descriptions such as ``\emph{Unspecified vulnerability in [..] allows local users to affect availability via vectors related to Kernel}'',\footnote{\url{https://web.nvd.nist.gov/view/vuln/detail?vulnId=CVE-2016-5469}} to more technically detailed information~\cite{Christey-2013-BHUSA}. 
Whereas the type of information one can gather generally covers type of vulnerability, attack procedure, and existence of threats~\cite{holm2015expert}, no guidance currently exists on the mapping of which information should be considered when performing an assessment over the \CVSS\ metrics. For example, analysing the attack procedure may provide details on the position of the attacker w.r.t. the vulnerable software component (captured by the \texttt{Attack Vector} \CVSS\ metric), but may not reveal useful information to evaluate which privileges are required to exploit the vulnerability (captured by the \texttt{Privileges Required} metric). 
This prevents the development and use of automatic tools that can provide useful summaries of available information that the assessor can use when performing his or her \CVSS\ evaluation of the vulnerability.


In this study we evaluate which information cues can aid the vulnerability assessment process as guided by the \CVSS\ standard, and should therefore be readily provided to  the assessor. {This paper's contributions can be summarized as follows:}
\begin{enumerate}
 \item Following guidelines from current standards~\cite{iso29147} and recent literature~\cite{holm2015expert,Roschke2009}, we identify four information categories over which vulnerabilities are described: \texttt{Assets}~\cite{Stoneburner:2002:SRM:2206240}, \texttt{Attack}~\cite{Roschke2009}, \texttt{Vulnerability type}~\cite{iso29147}, and \texttt{Known threats}~\cite{holm2015expert}.

 \item Building on recent research on the automatic identification of `requirement smells'
~\cite{Femmer2016}, we evaluate the number of \emph{information cues} (i.e.~phrases consisting of one or more words) associated with each of the identified information categories, and their affect on assessment error.

 \item We ask 67 students to score a set of 16 vulnerabilities using \CVSS. To evaluate the effect of different information cues, we devise a block experiment design in which each student is assigned randomly to a treatment\footnote{Treatments integrate baseline vulnerability descriptions with information provided by the standard body for \CVSS.} group, and compare assessment errors to identify which information cues are effective in aiding the final assessment and which are not.

\end{enumerate}

This paper unfolds as follows: Section~\ref{sec:background} discusses related work. Section~\ref{sec:method} outlines our research goal and questions, experiment setup, metrics, and hypotheses. We then presents our results (Section~\ref{sec:results}) and discuss their implications (Section~\ref{sec:discussion}). Section~\ref{sec:validity} and~\ref{sec:conclusions} discusses threats to validity and conclude.

\section{Background and Related Work}
\label{sec:background}
In security engineering controlled experiments have been performed to measure the effectiveness and efficiency of vulnerability analysis techniques and applications \cite{Scandariato2013,Allodi-2014-TISSEC,Allodi-13-IWCC}, security patterns in helping software designers \cite{Yskout2015}, and the application of different security methods for risk assessment \cite{labunets2013experimental}. 

Similarly, several authors studied the relation between vulnerability measures and risk scenarios. 
The operative aspects integrating security measures in production environments have been studied, among others, by 
Dashevskyi et al.~\cite{dashevskyi2016security} (who investigate settings where vulnerabilities are included in third party components),  
Zhang et al.~\cite{zhang2013predicting} (who predict bug fixing times by employing a Markov model based on field data), 
Zimmermann et al.~\cite{zimmermann2010makes} (who investigate the discrepancies between user-supplied bug information and information needed by the developers), and Zhao et al.~\cite{zhao2016discussions} (who evaluate the effect of early discussion on bug fixing). We integrate these findings by focusing on vulnerability information and evaluating which information aids the vulnerability fixing process.


Proposed measures for the identification of vulnerabilities in code rely on features of code such as code complexity and code churn~\cite{SHIN-etal-10-TSE}, whereas other authors propose keyword-based text-mining procedures to forecast vulnerabilities~\cite{walden2014predicting}. Thompson et al.~\cite{Thompson:2016:SDU:2901739.2901779} investigated the cognitive effort spent when breaking down software engineering tasks such as bug fixing. 
To aid a correct understanding of software requirements, natural language processing techniques such as keyword extraction have been used to detect quality defects in natural language specification~\cite{Femmer2016}. 
Experimentation often relates to factors such as the correctness and the positive or negative tone of requirements~\cite{MFME15}, and grammatical features such as passive or active voice requirements. While our approach is similar, we detect information cues in vulnerability description text to associate it with assessment errors, as opposed to measuring `bad wording' in software requirements.

\subsection{The Common Vulnerability Scoring System}

The \CVSS\ framework specification is the worldwide standard for vulnerability assessment and has been drafted by the dedicated \emph{First.org} Special Interest Group (SIG). 
The \CVSS\ framework 
provides a number of dimensions over which a vulnerability is assessed based on 
available information on the vulnerability. These dimensions are classified into three groups, or metrics: Base Metric (captures technical characteristics of the vulnerability), Temporal Metric (captures vulnerable conditions that change in time), and Environmental Metric (captures conditions that change by deployment environments).
The Base Metric Group is by far the most commonly used in practice \cite{Naaliel-ISSRE-14,Houmb-2010-JSS} and is the one officially used to describe vulnerabilities in the NIST's \emph{National Vulnerability Database} (NVD)~\cite{NVD}.

The Base Score assessment is organized in two conceptually different groups of sub-metrics~\footnote{A third metric group, \texttt{Scope}, is not reported here for brevity as it is not used in this study.};
\emph{Exploitability} metrics reflect the means by which an attacker can deliver a 
successful attack, whereas
\emph{Impact} metrics provide an assessment of the consequences of a successful attack 
on the impacted system. 

Exploitability metrics under \CVSS\ v3 are measured over four dimensions: 
Attack Vector (\AV), Attack Complexity (\AC), Privileges Required (\PR) 
and User 
Interaction (\UI). 
Impact metrics in \CVSS\ v3 are measured over the triad Confidentiality, Integrity and Availability. 
Table \ref{tab:CVSSbasemetrics} provides a summary description of the \CVSS\ v3 Base metrics. Full reference can be found at the official \emph{First.org} specification documentation~\cite{first-2015-cvss3}.
\begin{table}[t]
\centering
\small
\caption{Summary description of \CVSS\ v3 Base metrics}
\label{tab:CVSSbasemetrics}
\begin{tabular}{c p{1.5cm} p{3.9cm} p{1.4cm}}\toprule
\multicolumn{4}{c}{\textbf{Exploitability metrics}} \\
ID & Metric & Description & Values\\\midrule
\AV & Attack \newline Vector & Reflects how remote the attacker can be, to deliver the attack against the vulnerable component. The more remote, the higher the score. & \texttt{Physical, 
Local, Adj. Net., Network.}\\
\AC & Attack  \newline  Complexity & Reflects the existence of conditions that are beyond
the attacker's control for the attack to be successful. & \texttt{High, Low.}\\
\PR& Privileges Required & Reflects the privileges the attacker need have
on the vulnerable system to exploit the vulnerable component. & \texttt{High, Low, None.}\\
\UI&User  \newline  Interaction & Reflects the need for user interaction to deliver a successful attack. & \texttt{Required, None.} \\
\midrule
\multicolumn{4}{c}{\textbf{Impact metrics}} \\
ID & Metric & Description & Values\\\midrule
\C & Confidentiality & Measures the impact to the confidentiality of information.  & \texttt{None, Low, High.} \\
\I&Integrity & Measures the impact to the integrity of information. & \texttt{None, Low, High.}\\
\A&Availability & Measures the impact to the availability of the impacted component. & \texttt{None, Low, High.}\\
\bottomrule
\end{tabular}
\end{table}

\subsection{Information categories for vulnerability measurement}
\label{sec:infocues}


We evaluate the effect of the following information categories on the accuracy of \CVSS\ assessments:

 {\texttt{Assets}.} Security assessment and management standards such as NIST 800-30 and Common Criteria~\cite{CC2009,Stoneburner:2002:SRM:2206240} define the concept of `asset' as key to correctly evaluate the severity of the vulnerability impact. Information in this category includes details on type of affected system (e.g. a server or a client) or the component affected by the vulnerability (e.g. an operating system or a virtual machine).

{\texttt{Attack}.} Expert interviews conducted by Holm et al.~\cite{holm2015expert}, alongside other studies \cite{fruhwirth2009improving}, identify information regarding attack procedures as important to conduct an accurate vulnerability assessment. Attack procedures describe the actions that an attacker must perform to exploit the vulnerability: for example, the attacker may need to launch a \emph{man-in-the-middle} attack, or inject code in a webpage.

{\texttt{Vulnerability type}.} ISO 29147~\cite{iso29147} conceptualizes vulnerability information as related to a description of the vulnerability and its impact. This includes information on the type of vulnerability and its causes in the program's code. For example, an erroneous bound checking of a memory array may lead to \emph{memory corruption} vulnerabilities; similarly, erroneous input validation on a web form may lead to \emph{cross-site-scripting} (XSS) vulnerabilities. 

 {\texttt{Known threat}.} Several studies \cite{barnum2005knowledge,Roschke2009,holm2015expert} suggest that information on existing threats should also be provided to aid a better vulnerability assessment. This information includes details on the existence of \emph{proof-of-concept} exploit (\emph{PoC}), active exploitation in the wild, or incidents linked to the specific vulnerability.

\section{Methodology}
\label{sec:method}
In this paper, following the discussion in Sec.~\ref{sec:infocues}, we investigate the following research question:
\emph{How does information on \{\texttt{Asset}, \texttt{Attack}, \texttt{Vuln. type}, \texttt{Known threat}\} impact assessment errors?}\\

\subsection{Experimental settings}

To address these four research questions, we perform an experiment where subjects are asked to score sixteen vulnerabilities using \CVSS. Each vulnerability is associated with its description from the National Vulnerability Database (NVD)\cite{NVD} and a treatment consisting of \emph{additional information} on the vulnerability (on top of its baseline NVD description) provided by the \CVSS\ SIG~\cite{first-2015-cvss3-exampleDoc}.
Table~\ref{tab:exampleDescription}
\begin{table*}[t]
\centering
\small
\caption{Example of vulnerability descriptions and treatments given to the students.}
\label{tab:exampleDescription}
\begin{minipage}{0.95\textwidth}
Example of four CVE descriptions and treatments assigned to students. We obtained this descriptions from the NVD. Treatment are obtained from the official CVSSv3 example guide~\cite{first-2015-cvss3-exampleDoc}. The column `Treatment effect' outlines the effect on error rate of the treatment. Indicated p-values are Holm-corrected for multiple comparisons over \CVSS\ metrics. We highlighted in bold relevant excerpts that explain the treatment effect.  Significance of the treatment effect is evaluated with a Wilcoxon rank-sum test.
 \vspace{0.05in}
\end{minipage}
\begin{tabular}{l p{6cm} p{5.7cm} p{2.5cm}}
\toprule
CVE & NVD Description & Treatment & Treatment effect\\\midrule
 

 CVE-2014-3566  &The SSL protocol 3.0, as used in OpenSSL through 1.0.1i and other products, uses nondeterministic CBC padding, which makes it easier for man in the middle attackers to obtain plaintext data via a padding- oracle attack, aka the "POODLE" issue. &  A typical treatment is that a \textbf{victim has visited a web server and her web browser now contains a cookie} that an attacke wishes to steal. For a successful attack, the \textbf{attacker must be able to modify network traffic between the victim and this web server}, and both victim and system must be willing to use SSL 3.0 for encryption. &  


 Decrease error on \AC\ ($p<0.10$)

  Decrease error on \UI\ ($p<0.01$)\\
  \\

 CVE-2012-0384 & Cisco IOS 12.2 through 12.4 and [..]
 before 3.2.2SG, when AAA authorization is enabled, allow remote authenticated users to bypass intended access restrictions and execute commands via a (1) HTTP or (2) HTTPS session, aka Bug ID CSCtr91106.  & \textbf{This vulnerability is post authentication on the administrative interface of the Cisco device.} Therefore to attack a typical installation, the attacker would need access to the trusted / internal side of the IOS.  & Increase error on \PR\ ($p<0.01$) \\

\bottomrule
\end{tabular}
\end{table*}
reports example vulnerability descriptions and treatments used for the experiment. The column `Treatment effect' reports the effect of the treatment on the accuracy of the assessment, which is discussed in detail in Section~\ref{sec:results}.

 Subjects were given 90 minutes to complete the assessment irrespective of the treatment selection. Hence, each subject had on average about 6 minutes per vulnerability. In accordance with literature on the subject \cite{pennington2007effects}, the time was selected on the basis of previous trial experiments previously conducted in similar settings. 

 \subsection{Vulnerabilities and Subject Selection}
The sixteen vulnerabilities employed in the experiment are obtained from the official CVSS v3 Example document drafted by the First.org SIG for CVSS~\cite{first-2015-cvss3-exampleDoc}. The vulnerabilities included by the SIG have been chosen to represent the full set of CVSS metrics, and are actively 
used for training purposes by members of the SIG consortium within the respective organizations.  Each vulnerability in the document is associated with its official public description from the National Vulnerability Database~\cite{NVD} and additional information added by the \CVSS\ SIG. 
The subjects of this study are 67 students enrolled in the software engineering study program, who registered for a software security course.

\subsection{Measures}
\label{sec:measures}

\paragraph{Information cues}  To quantify the amount of information in a vulnerability description (for each information category identified in Sec.~\ref{sec:infocues}: \texttt{Asset, Attack, Vulnerability type, Known threat}) 
we employ a methodology originally developed to automatically identify `smells' in software requirement specifications~\cite{Femmer2016}. The original methodology employs keyword-matching to identify standard-defined criteria for quality of requirements in the analysed text. As no such standard exists for software vulnerability descriptions, in our study we identify keywords relevant to each of the identified information categories by manually analysing over 100 randomly sampled vulnerabilities from NVD. Keywords are  selected as indicators of what information is present in the description. For example, the keyword `\emph{remote attacker}' indicates that the vulnerability description explicitly reports information relative to the information category \texttt{Attacker}. 
Information cues are measured as the number of keyword matches in a baseline vulnerability description and in the corresponding treatment.
Table~\ref{tab:definitions} reports a sample of the keywords identified for each information category.  The full keyword list is available in the online appendix.\footnote{\url{https://github.com/tum-i22/information-cues}}
 \begin{table*}[t]
 \centering
 \small
 \caption{Definitions of information categories and selection of respective keywords.}
 \label{tab:definitions}
 \begin{tabular}{p{.11\textwidth} p{.2\textwidth} p{.1\textwidth} p{.5\textwidth}}
 \toprule
 Information categories & Definition & Reference & Keywords  \\
 \midrule
   \texttt{Asset} & Assets are entities that users or vendors value and contain vulnerabilities.  & \cite{first-2015-cvss3,Stoneburner:2002:SRM:2206240,iso27001} & 
   hardware, 
guest virtual machine, 
host, 
vm, 
device, 
client, 
server, 
operating system, 
version, 
product, 
affected version, 
affected product, 
vulnerable, 
vulnerable software, 
vulnerable hardware, 
affected software, 
affected hardware, 
software
   \\
\texttt{Attack} &  Actions and entities that can adversely act on assets by exploiting vulnerabilities. & \cite{holm2015expert,Roschke2009,barnum2005knowledge} & 
attacker, 
malicious user, 
remote authenticated user, 
remote user, 
man in the middle, 
unauthenticated remote attacker, 
spoofing, 
inject code, 
manipulate pointers, 
cache poisoning, 
open malicious file, 
birthday attack 
 \\
   \texttt{Vulnerability type}  &  Describes the technical flaws that can be exploited and the impact of the exploitation.  & \cite{iso29147} & 
improper bounds checking, 
insufficient randomness, 
memory corruption, 
buffer overflow, 
cross-site scripting, 
broken authentication, 
insecure cryptographic storage, 
failure to restrict URL access,
cross-site request forgery (CSRF)
 \\
        
 \texttt{Known threat}  &  Describe known threats that can exploit the vulnerability  & \cite{holm2015expert,barnum2005knowledge} & 
 known threats, 
threat, 
known attacks, 
information about known threats, 
exploit, 
proof-of-concept, 
incident activity, 
incident, 
known incident\\ 
  \bottomrule
 \end{tabular}
 \end{table*}


\paragraph{Assessment errors} To evaluate assessment errors, we compare the subjects' \CVSS\ assessments on the vulnerabilities with those performed by the \CVSS\ SIG. In this study we do not consider magnitude or directionality of error, but only the presence of a correct ($error=0$) or wrong ($error=1$) assessment for each \CVSS\ metric (\AV, \AC, \UI, \PR, \C, \I, \A), for all vulnerabilities.

\paragraph{Subject characteristics} Each subject was asked to complete a background questionnaire. We collected data relative to: security expertise of the student; software engineering expertise; years of prior work experience; years of enrollment in a Computer Science major; university courses completed. Students where asked to perform both a self-assessment on their expertise and to answer a set of multiple-choice technical questions on relevant areas of software security and engineering. Each technical question has only one correct answer. The questionnaire is available in the online appendix. Results are discussed in Section~\ref{sec:subjects}. 

\subsection{Hypotheses}
\label{sec:hyp}


\paragraph{\texttt{Asset}} Because \texttt{Asset} provides information regarding the target of the attack (e.g. a browser, or a server) we expect this information category to reduce error assessments on the impact metrics \C, \I\ and \A. For example, an attack on a browser may violate the Confidentiality of information stored in cookies or browsing history, whereas an attack on a server may affect the service Availability. We formulate the following \emph{null} hypothesis:

\begin{hypothesis}
\label{hyp:assets}
$H_0$: The \texttt{Asset} information category does not \emph{reduce} error rates for the \C, \I, \A\ metrics.
\end{hypothesis}

\paragraph{\texttt{Attack}} Information on \texttt{Attack} adds details on the actions that the attacker has to perform to exploit the vulnerability. Therefore, we expect this information category to reduce assessment error for the \AV\ metric (position of the attacker with respect to the vulnerable component), and the \AC\ metric (reflecting conditions outside of the attacker control). Additionally, indications on the attacker actions may give significant indications for the impact of the vulnerability. For example, performing a \emph{cache poisoning attack}\footnote{A cache poisoning attack requires the attacker to modify some cached record (e.g. a DNS response) such that at the next request the victim will receive the counterfeit information added by the attacker. This may lead to spoofing attacks with possible losses on at least Confidentiality and Integrity.} has clear repercussions on \C\ and \I. Denial of service attacks may indicate losses on \A. We formulate the following \emph{null} hypothesis:

\begin{hypothesis}
\label{hyp:attack}
$H_0$: The \texttt{Attack} information category does not \emph{reduce} error rates for the \AV, \AC, \C, \I, \A\ metrics.
\end{hypothesis}

\paragraph{\texttt{Vulnerability type}} Information on \texttt{Vulnerability type} provides information on the complexity of an attack, e.g. by specifying that the vulnerability is due to insufficient randomness in a specific variable. Information regarding specific vulnerability types (e.g. cross-site-scripting vulnerabilities) and required authentication levels give information on \PR\ and \UI. We formulate the following \emph{null} hypothesis:

\begin{hypothesis}
\label{hyp:vtype}
$H_0$: The \texttt{Vulnerability type} information category does not \emph{reduce} error rates for the \AC, \PR, \UI\ metrics.
\end{hypothesis}

\paragraph{\texttt{Known threat}} From the \CVSS\ specification, the Base Metric should only consider information relative to the technical characteristics of the vulnerability. Specifically, \texttt{Known threat} information may be relevant in subsequent assessments to evaluate risk of attack (e.g. involving the \CVSS\ temporal metrics~\cite{first-2015-cvss3}), but may confuse the baseline assessment of the vulnerability. For example, \texttt{Known threat} information may increase error on \AC\ as the existence of known threats may suggest that the vulnerability can be easily exploited, e.g. building up on the existing PoC. Similarly, information on known attacks may influence impact assessments to reflect those of the known incidents.
Therefore, we expect \texttt{Known threat} to be generally detrimental to the assessment of \AC\ and \C,\I,\A. We formulate the following \emph{null} hypothesis:

\begin{hypothesis}
\label{hyp:threats}
 $H_0$: The \texttt{Known threat} information category does not \emph{increase} error rates for the \AC, \C, \I, \A\ metrics.
\end{hypothesis}

\subsection{Experimental procedure}

Before the experiment subjects were given a lecture on vulnerability assessment with \CVSS. The lecture covered all aspects of the standard required for the experiment. With the objective of increasing subject's confidence in the procedure, a demo session scoring five vulnerabilities from the \CVSS\ documentation (not included in the experiment) was performed during the lecture.

Subjects were given a handout with the official \CVSS\ specification, and a printout spreadsheet containing the sixteen vulnerability descriptions. 
Subjects were randomly assigned to a treatment group and received
additional information on each vulnerability together with the NVD description. Subjects had to 1) complete the questionnaire described in Sec.~\ref{sec:measures}; 2) read each vulnerability description; 3) indicate which value for each of the \CVSS\ metrics in Tab.~\ref{tab:CVSSbasemetrics} better reflect the vulnerability description.

\subsection{Analysis procedure} To test our hypotheses
we employ a set of multilevel mixed effect regression models of the form:
$y^{m}_{s,c} = \textbf{Z}_{s}\boldsymbol\beta_1 + \textbf{X}_{c}\boldsymbol\beta_2  +  u_{s} + v_{c} + \epsilon^{m}_{s,c}$,
where $y^m_{s,c}$ reflects the presence or absence ($y^m_{s,c} \in \{1,0\}$) of an assessment error on the metric $m\in\{\AV,\AC,\UI,\PR,\C,\I,\A\}$ by student $s$,  over vulnerability $c$; $\textbf{Z}_s$ is the control vector of  subject characteristics, and $\textbf{X}_{c}$ is the vector of information cues for each category measured on vulnerability $c$. The remaining terms account for random effects for the first level in the hierarchy, students ($u_{s}$); and the second, vulnerability ($v_c$). Each hypothesis is evaluated in accordance with the respective coefficient sign and its significance.


\section{Results}
\label{sec:results}
\subsection{Overview of subjects}
\label{sec:subjects}

 Before executing the experiment, we asked students to fill out a questionnaire that provides an overview of their background (twenty multiple-choice questions) and relevant security and software engineering expertise (six self-assessment questions and six technical questions). All questions where divided in \emph{security} and \emph{software engineering} questions.
From the 67 participants, 14 were Bachelor students, the rest were Master students. 
36\% of the participants have part-time work occupations. 
 Looking at the techninal security questions, the mean score was 0.57, with 1 indicating all correct answers and 0 no correct answer. The standard deviation is relatively small at 0.27 points. Similar scores are identified for the software engineering technical questions. 

\subsection{Illustrative analysis example}
\label{sec:initialanalysis}
 Table~\ref{tab:exampleDescription} reports two example vulnerability descriptions for which treatments have a significant effect on assessment errors for at least one \CVSS\ metric. We report one vulnerabilities where we observe negative effects on the error (\texttt{CVE-2014-3566}), and one where we observe positive effects (\texttt{CVE-2012-0384}). Information that explains the difference is highlighted in bold in the Table. In the following, the correct metric assessment is reported next to the \texttt{CVE} vulnerability identifier:


\textbf{{\texttt{CVE-2014-3566}} (\AC:\texttt{High}, \UI:\texttt{Required})}. Students that received the treatment were less likely to err at identifying: (a) conditions outside of the attacker control ($p<0.10$), as the treatment specifies that ``\emph{[the] attacker must be able to modify network traffic between the victim and this web server}'', suggesting a man-in-the-middle condition; (b) the requirement on \UI, specifying that the attack is possible only after ``\emph{[the victim] has visited a web server}'' ($p<0.01$).

\textbf{{\texttt{CVE-2012-0384}} (\PR:\texttt{Low})}. The treatment significantly increases chances of error over \PR\ ($p<0.01$). The treatment states that to trigger the vulnerability \emph{``the attacker would need access to the trusted / internal side of the IOS.''}. Any user authenticated in the network would be able to access the interface (i.e. only non-privileged authentication to the network is required). However, the additional information that \emph{``the vulnerability is post authentication on the administrative interface of the Cisco device''}, can be misleading in that the attacker does not need to be logged in the administrative panel, but only capable of reaching it from the network (in which he/she must be authenticated).

In our examples, additional information could either aid or hinder a correct assessment by, for example, misleading wording of relevant information (e.g.CVE-2012-0384): in accordance with previous findings in sw engineering~\cite{eppler2004concept,pennington2007effects}, both \emph{quantity} and \emph{quality} of information may affect task execution. Unfortunately, neither can be a realistic requirement for an informative vulnerability description as they do not provide a clear guidance on \emph{which} information cues should be provided. 

\subsection{Tratment effect on assessment error}
\label{sec:regression}

To identify the effect of the measured information cues we employ a set of mixed-effect regression analyses. For the model selection we relied on the Akaike Information Criterion\footnote{Considered control variables: $Z1:$ security expertise of the student; $Z2:$ software engineering expertise; $Z3:$ work experience; $Z4:$ years of enrollment in a Computer Science major $Z5:$ university courses completed.}. The only significant student characteristic is \emph{security expertise} ($E^{sec}$). Correlation between the independent variables is always below 0.2.

We first check for the possible correlation between length of vulnerability description (expressed as word counts) and error rates, and find that neither the length of the original NVD description nor the length of the treatment text have significant effects on the observed error. We therefore proceed with the analysis of the effect of the information cues.

For our final regression, the regressors are count of information cues measured in the original $\texttt{NVD}$ description and those added by the assigned treatments $T$. All variables are standardized.  The final regression equation over the binomial response variable representing assessment error $\hat y= y^{m}_{s,c}$ is:

\begin{equation}
\nonumber
\hat y_i = \beta_0+\beta_1 E^{\text{sec}}_i + \boldsymbol\beta_2 \text{\textbf{Cues}}^\texttt{NVD}_{i} + \boldsymbol\beta_3 \text{\textbf{Cues}}^\texttt{T}_{i} + u_s + v_c + \epsilon_i
\end{equation}

Results are reported in Table~\ref{tab:regressionResults}.
\begin{table*}[t]
\caption{Regression results}
\label{tab:regressionResults}
\centering
\small
\begin{minipage}{0.85\textwidth}
\footnotesize
Regression results for our equations.
p-values for the fixed effects are computed by using Satterthwate's estimation for degrees of freedom as provided by the R package \texttt{lmerTest}. Standard errors are indicated in parenthesis.
Regression coefficients are reported for the information cues all students received (as provided in the original NVD description of the vulnerability) and for the additional information cues included in the treatment. All variables are standardized. The original NVD descriptions do not have any information regarding \texttt{Known threats}, which is therefore only relevant for the provided treatments. 
 An anova test of variance indicates that the intercepts for students and CVEs significantly vary between subjects and vulnerabilities. \vspace{0.05in}
\end{minipage}
\begin{tabular}{p{5cm}rrrrrrr}
\toprule
&\multirow{2}{*}{model \AV}&\multirow{2}{*}{model \AC}&\multirow{2}{*}{model \UI}&\multirow{2}{*}{model \PR}&\multirow{2}{*}{model \CI}&\multirow{2}{*}{model \II}&\multirow{2}{*}{model \AI}\\

\emph{Fixed effects}\\
\midrule
(Intercept)&-0.804$^{***}$&-0.667$^{***}$&-2.035$^{***}$&-1.008$^{***}$&0.504$^{**}$&0.088&-0.511\\
&(0.203)&(0.200)&(0.376)&(0.331)&(0.224)&(0.246)&(0.312)\\
$E^{sec}$&-0.088&-0.007&-0.233$^{**}$&-0.137&-0.190$^{*}$&-0.246$^{***}$&-0.080\\
&(0.081)&(0.098)&(0.110)&(0.128)&(0.100)&(0.093)&(0.095)\\
\multicolumn{1}{p{5cm}}{\emph{Information cues from original description}}\\\cmidrule(lr){1-1}
\texttt{Assets}&0.384$^{*}$&0.097&-0.550&0.246&-0.035&0.314&-0.696$^{**}$\\
&(0.198)&(0.192)&(0.383)&(0.328)&(0.221)&(0.232)&(0.320)\\
\texttt{Attack}&-0.555$^{***}$&0.062&0.111&-0.209&-0.013&-0.034&0.388\\
&(0.211)&(0.196)&(0.386)&(0.333)&(0.228)&(0.238)&(0.327)\\
\texttt{Vulnerability type}&0.085&-0.330$^{**}$&-0.362&-0.530$^{***}$&0.169&0.032&-0.150\\
&(0.149)&(0.135)&(0.228)&(0.166)&(0.143)&(0.139)&(0.163)\\

\multicolumn{1}{p{5cm}}{\emph{Additional information cues from treatment}}\\\cmidrule(lr){1-1}
\texttt{Assets}&-0.191&0.169&-0.282&-0.054&0.121&-0.069&-0.100\\
&(0.129)&(0.123)&(0.175)&(0.131)&(0.113)&(0.125)&(0.108)\\
\texttt{Attack}&-0.036&-0.280$^{**}$&-0.238&-0.031&-0.313$^{***}$&-0.202$^{*}$&-0.183\\
&(0.116)&(0.127)&(0.183)&(0.139)&(0.112)&(0.114)&(0.120)\\
\texttt{Vulnerability type}&-0.108&-0.067&-0.563$^{***}$&-0.103&-0.015&-0.072&-0.052\\
&(0.098)&(0.116)&(0.171)&(0.125)&(0.100)&(0.100)&(0.112)\\
\texttt{Known threats}&0.176&0.528$^{***}$&0.494$^{***}$&-0.206&0.325$^{***}$&0.017&0.278$^{**}$\\
&(0.128)&(0.138)&(0.177)&(0.138)&(0.121)&(0.127)&(0.131)\\
  \emph{Variance of random intercepts}&\\
\midrule
Student&0.045&0.263&0.101&0.637&0.314&0.210&0.200\\
       CVE&0.464&0.436&1.821&1.473&0.636&0.706&1.400\\
Pseudo-$R^2$ (Fixed effect) & 0.09 & 0.09 & 0.14 & 0.09 & 0.03 & 0.05 &0.09\\
Pseudo-$R^2$ (Fixed and random eff.) & 0.22 & 0.25 & 0.46 & 0.44 & 0.25 & 0.26 &0.39\\
\bottomrule
\end{tabular}
\begin{minipage}{0.85\textwidth}
\small
\vspace{0.05in}
Signif. codes:  `***' 0.001; `**' 0.01; `*' 0.05; `.' 0.1.
\end{minipage}
\end{table*}
A negative, significant coefficient indicates a decrease in the chances of error. Positive, significant coefficients indicate an increase in chances of error. Security expertise tends to reduce error although it is not a significant factor for all metrics. Overall, we find consistent estimations for each information category. In general, information cues on \texttt{Attack} and \texttt{Vulnerability type} aid the scoring for all metrics. \texttt{Asset} creates mixed results, whereas \texttt{Known threat} is always counter-productive. The fixed effects account for about 10\% of the overall variance in the model across all metrics, with only a few exceptions in either direction (14\% for \UI, 3\% for \CI). The inclusion of the random effects accounts for in between 22\% and  46\% of the variance, indicating a good overall fit.

\paragraph{{RQ1: How does information category \texttt{Asset} impact assessment errors?}} Error rates on \A\ are negatively impacted by this information category; for example, if the vulnerable asset is a server, service availability can be likely compromised by an attack. Additionally, we find that information on \texttt{Assets} \emph{increase} the error on the \AV\ metric, albeit the effect is only weakly significant. Some assets (e.g. a router or a server) may be correlated with \AV:\texttt{Network} assessments, whereas in specific cases the attacker may need be locally authenticated on the asset. We provide two examples of this from our experiment in the next Section. We reject the null hypothesis of Hyp.~\ref{hyp:assets} for \A\ and accept the alternative that there is a decrease in error. We do not reject the null for \C,\I.

\paragraph{{RQ2: How does information category \texttt{Attack} impact assessment errors?}} This information category improves accuracy on \AV, as it can clearly indicate the position of the attacker. Similarly, we find a negative effect on error for \AC. For example, a \emph{man-in-the-middle} attack suggests a high condition for this metric~\cite{first-2015-cvss3}.
For the \CVSS\ impact triad \C\I\A, information regarding the attack decreases error for Confidentiality and Integrity. For example, a cache poisoning attack implies an impact on the integrity of the cached information. We reject the null hypothesis of Hyp.~\ref{hyp:attack} for \AV,\AC,\C,\I\ and accept the alternative that there is a decrease in error. We do not reject the null for \A.

\paragraph{RQ3: How does information category \texttt{Vulnerability type} impact assessment errors?} For \AC, information on the type of vulnerability favours assessment accuracy.  For example, specifying that the vulnerability is caused by insufficient randomness (e.g. of a hash function) may indicate that the attacker will typically have to find a collision before actively exploiting the vulnerability. \texttt{Vulnerability type} also reduces chances of error on \UI. For example, a cross-site-scripting vulnerability typically requires the user to click on a malicious link. The effect on \PR\ is similar: this information cue may clarify whether some level of privilege is required to launch the attack. For example, privilege escalation vulnerabilities typically require some level of authentication. We reject the null hypothesis of Hyp.~\ref{hyp:vtype} for \AC,\PR,\UI, and accept the alternative that there is a decrease in error.

\paragraph{{RQ4: How does information category \texttt{Threat} impact assessment errors?}} In general, we find that this information cue increases the chances of error. From the \CVSS\ specification, the Base Metric should only consider information relative to the technical characteristics of the vulnerability. Hence, the existence of an exploit or of a demonstrated attack is unnecessary information that need be processed by the assessor. For example, information on the existence of a demonstrated attack may increase the error on \AC, as previously discussed (cf.~\ref{sec:hyp}). Similar considerations can be made for the other metrics. We reject the null hypothesis of Hyp.~\ref{hyp:threats} for \AC,\C,\A, and accept the alternative that there is an increase in error. We do not reject the null for \I.
\begin{table}[h!]
\centering
\small
\centering
\begin{tabular}{l l l l l}
\toprule
 Hyp. & Inf. Cue  & \multicolumn{2}{c}{$H_0$}& \multicolumn{1}{c}{$H_1$}\\
\cmidrule(lr){3-4}
&& Reject & No reject&\\
\midrule
Hyp.~\ref{hyp:assets} &\texttt{Assets}  & \A & \C,\I &  Err. Decrease\\
Hyp.~\ref{hyp:attack} &\texttt{Attack}  & \AV,\AC,\C,\I & \A&  Err. Decrease\\
Hyp.~\ref{hyp:vtype} & \texttt{Vuln. Type}  & \AC,\PR,\UI & - &   Err. Decrease\\
Hyp.~\ref{hyp:threats} & \texttt{Known threats} & \AC,\C,\A & \I &  Err. Increase\\
\bottomrule
\end{tabular}
\end{table}


\section{Discussion}
\label{sec:discussion}





Our results indicate that `baseline' vulnerability descriptions can be significantly improved by including additional information. Information of type \texttt{Attack} and \texttt{Vulnerability type} are particularly effective in increasing the accuracy of vulnerability assessment by reducing error on the whole set of Exploitability metrics (cf. Table~\ref{tab:CVSSbasemetrics}) \AV, \AC, \UI, \PR. 


In our sample, 
additional information on attacker actions significantly decrease error on \AC, indicating that additional information on \texttt{Attack} was missing from the original text. Similarly, the \texttt{Attack} information added by our treatments also significantly decrease the error for \C\ and \I. Our results suggest  that security expertise helps interpreting this information (e.g. a `cache poisoning' attack). 

The information on \texttt{Vulnerability type} conveyed by standard vulnerability descriptions seem not to be significantly improved by our treatment for \AC\ and \PR, whereas there is a highly significant improvement in assessment accuracy for \UI. Information regarding the type of vulnerability such as a file-based buffer overflow or a cross-site-scripting 
vulnerability should be included in vulnerability descriptions. This is again in accordance with the negative significant coefficient for $E^{sec}$, indicating that security expertise is significant in correctly understanding the type of vulnerability. 

An interesting finding is that \texttt{Asset} contribute in \emph{increasing} error on \AV. Certain information on \texttt{Asset} may correlate with certain \AV\ values; for example, if the vulnerable asset is a server, \AV:\texttt{Network} assessments may be more likely.
For example, $42\%$ of the students erroneously classified \texttt{CVE-2014-6271} as \AV:\texttt{Local}, likely as the vulnerability is specified, in the original NVD description, to affect ``\emph{GNU Bash through 4.3 [..] [in] situations in which setting the environment occurs across a privilege boundary from Bash execution}''. Here the vulnerable asset is clearly the GNU Bash, which may suggest that the user need be authenticated locally to reach the vulnerability. However, in the worst case this is possible without any local access to the environment, as specified in the description: ``\emph{the vulnerability can be exploited by [..] mod\_cgid modules in the Apache HTTP Server, scripts executed by unspecified DHCP clients [..]}'', which indicates a \texttt{Network} vector for the attack.
Similarly, in our sample, 89\% of the students erroneously categorized \texttt{CVE-2012-1516} as \AV:\texttt{Local}. The description reports that \emph{``it is possible to manipulate data pointers within the Virtual Machine Executable (VMX) process''}, which suggests that the user need be locally authenticated on the machine to access the process. This is not a condition for \AV:\texttt{Local} as the vulnerability can be reached by the \emph{``handler function for RPC commands''}~\cite{first-2015-cvss3-exampleDoc}, a procedure to send remote commands to a process.
Both examples suggest that a more precise definition of  \texttt{Attacker} actions may contribute in decreasing the effect.

Finally, information on \texttt{Known threats} is regarded by security experts as of primary importance to assess vulnerability risk~\cite{holm2015expert}. However, we find first evidence that it consistently increases chances of error, as the \CVSS\ Base metric should only consider technical details on the vulnerability. 

Following Devanbu et al.'s recommendations on the impact of empirical findings on software practices~\cite{Devanbu:2016:BEE:2884781.2884812}, we further discuss practical implications of this work.

\textbf{Implications for vulnerability communication.} Our results suggest that baseline vulnerability descriptions contain only a limited set of the information that leads to an accurate \CVSS\ assessment. Additional information on \texttt{Attack}, \texttt{Vulnerability type}, and \texttt{Assets} may result in more informative vulnerability descriptions. Following our results, standards and best practices for vulnerability communication, including \CVSS\ itself, may provide guidelines for the communication of informative vulnerability descriptions~\cite{iso29147}. Our results suggest that inclusion of information of the \texttt{Threat} category should be discouraged. Further, our results identify dimensions over which vulnerability information can be automatically categorized and provided to vulnerability assessors.

\textbf{Implications for software security practices.}
Our findings can help practitioners in identifying information that is significant for a vulnerability assessment over each specific metric~\cite{pennington2007effects}. For example, the assessor performing an evaluation of the \AV\ metric may look specifically for \texttt{Attack} information. Similar considerations can be made for the other metrics (cf. Table~\ref{tab:regressionResults}). Additionally, the assessor should deliberately ignore any information on \texttt{Known threats}, if present. By replicating this work, it could be possible to build `confidence intervals' around vulnerability assessment that account for errors in the estimate. These intervals could then be accounted for when prioritizing vulnerability fixing.



\section{Threats to validity}
\label{sec:validity}

\emph{Conclusion validity.} To avoid introducing noise in the vulnerability descriptions and treatments used in the experiment, all descriptions and treatments have been chosen from the official documentation released by the the First.org standardisation team, used for official training for the standard. 

\emph{Internal validity.} Results may be confounded by order of treatment or learning effects. As we can not cover all treatment combinations, our experiment design is not full-factorial. However, we accounted for all combinations of treatments for similar vulnerabilities that might confound results. The identification of our keywords for the measurements of information cues in the vulnerability descriptions was performed independently by three authors of the paper.
To minimize chances of bias, the experiment was performed before the final example documentation was publicly released.

\emph{External validity.} Following \cite{host2000using} we consider students suitable subjects for relative performance measures. Students were informed that the exercise is not graded. All received the same training on the \CVSS\ scoring system at the beginning of our experiment. We controlled for potentially relevant characteristics of our subjects, including security expertise and work experience. 

\section{Conclusions}
\label{sec:conclusions}

In this paper we investigate which information cues aid vulnerability assessment by humans. 
We based our definition of relevant information on current standards and best practices~\cite{iso29147,Stoneburner:2002:SRM:2206240}, and recent research findings by other authors~\cite{holm2015expert,Roschke2009}. 
Our results provide first indication that, in general, additional information cues on \texttt{Asset}, \texttt{Attack}, and \texttt{Vulnerability type} on top of the baseline vulnerability descriptions may aid the assessment process, whereas information cues on \texttt{Threat} hinders it.

An interesting venue for future research is to explicitly consider the effect of information security knowledge by devising experiments with security professionals. 
Additionally, this work opens toward research considering measures of complexity to evaluate whether there exist boundaries over which the cognitive performance of the assessor decays.

\section{Acknowledgments}
This project has been partly supported 
by the University of Trento, Italy, through the EIT Digital Master School 2016 - Security and Privacy programme, and by the \grantsponsor{NWO}{NWO}{https://www.nwo.nl} through the SpySpot project no. \grantnum{NWO}{628.001.004}. 

\bibliographystyle{ACM-Reference-Format}
\bibliography{short-names,security-commonb,biblio}

\end{document}